# ASSESSING THE FEASIBILITY OF DEVELOPING A FEDERATED ERP SYSTEM

Michael Gall[1], Thomas Grechenig[1] and Mogens Bjerre[2]

[1]Institute for Industrial Software, Technical University Vienna, Vienna, Austria
michael.gall@inso.tuwien.ac.at
[2]Copenhagen Business School, Solbjerg Plads, Frederiksberg, Denmark
mb.marktg@cbs.dk

## ABSTRACT

*In past years ERP Systems have become one of the main components within the corporate IT structure. Several problems exist around implementing and operating these systems within companies. In the literature one can find several studies about the problems arising during the implementation of an ERP system. The main problem areas are around the complexity of ERP systems. One vision to overcome some of these problems is federated ERP. Federated ERP systems are built of components from different vendors, which are distributed within a network. All components act as one single ERP system from the user perspective. The decreased complexity of such a system would require lower installation and maintenance cost. Additional, only the components which are needed to cover the company's business processes would be used. Several theories around this concept exist, but a feasibility assessment of developing a federated ERP system has not been done yet. Based on a literary analysis of existing methods for feasibility studies, this paper is applying strategic planning concepts and referential data from the traditional ERP development to provide a first assessment of the overall feasibility of developing a platform for federated ERP systems. An analytical hierarchical approach is used to define effort and effect related criteria and their domain values. The assessment as the criteria is done in comparison to the development of a classical ERP system. Using the developed criteria, a net present value calculation is done. The calculation of the net present value is done on an overall, not company specific level. In order to estimate the weighted average cost of capital, the values from successful software companies are used as a baseline. Additional potential risks and obstacles are identified for further clarification.*

## 1. INTRODUCTION

An ERP (Enterprise Resource Planning) application is an integrated business management software application package, which consists of multiple components and a central database. It is integrating various different functional areas within a company. Its benefits are the reduction of data redundancies and the possibility to map integrated business processes. [1], [2] The centralized data storage forms the architectural basis of modern ERP systems. Modern ERP systems consist of many software components which are interoperating with each other. The business logic is held in a central place, the application server. The interactions of the different components form a complex network of dependencies. Out of this complexity several problems arise. From a customer's point of view, the main problems are that not all the components are required in every installation [3] and that the usage, maintenance and customization of these systems are too expensive due to the complexity. [4], [5], [6] Additionally ERP manufacturers are facing problems on how to effectively develop new features in such a system as for example the testability of the system is one of the major problems. [7] In recent years, ERP system manufacturers have started to focus on the small and midsize market. One reason for this is a saturation of the large enterprise market, because larger organizations have already implemented an ERP solution. Additionally, an increasing need for the integration of systems across company borders provides extended possibilities in this market segment. [8] The Western European ERP market for companies in the size band of 1 to 99 employees is approx. 1 billion euro and shows a growth of 7 %. [3] So this market is providing a great opportunity for ERP manufacturers. But in this market segment, small and medium size businesses are unable to cover the cost of ERP systems. Especially the maintainability and customizability of complex monolithic becomes a problem in this market segment. [3], [9] This indicates a necessity of solving the problem that arises from traditional ERP systems. A different approach to monolithic ERP systems is the concept of federated ERP. [10], [11], [12], [13], [14], [15], [16], [17] A federated ERP system (FERP system) is an ERP system where the individual system components are distributed within a computer network. From a user point of view, they appear as a single ERP system. Only the required components will be used and installed. The individual components can be developed and Different ERP system components can be developed by different vendors. [12], [13], [14] Currently these theories are in the early stages, without concrete implementation plans. In order to provide a first assessment of the feasibility for building such a system, a criteria catalogue regarding effort and effect will be used.

## 2. LITERATURE REVIEW

Various sources deal with ERP implementation and post implementation issues and their impact on enterprises. [8], [18], [19], [20], [21] These studies point out among others manageability and total cost of ownership as pain points. In the recent years, the focus of the ERP market has fallen on the small and medium size enterprises. One reason for this is a saturation of the large enterprise market, because larger organizations have already implemented an ERP solution. Additionally, an increasing need for the integration of systems across company borders provides extended possibilities in the lower market segment. [8] The Western European ERP market for companies in the size band of 1 to 99 employees is approx. 1 billion euro and shows a growth of 7 % per annum. [3] This indicates a necessity of mitigating the problems arising in the implementation and post implementation phase of an ERP project. The concept of federated ERP systems mitigates some of the observed issues. In existing ERP literature Brehm and Marx Gomez explore the concept and architecture with regard to federated ERP systems. Various papers exist covering the technical [10], [11], [12], [13], [14], [16] and architectural [10], [11], [15], [17] aspects of federated ERP systems. So far, only limited research has been done regarding concrete development of such a system. Only one paper deals with the business model which could be used for the development of such systems. [22] From an overall strategic point of view the feasibility of developing such a system has not been discussed yet. A first approach towards this topic should be given in this paper.

## 3. RESEARCH METHODS

This paper is part of an on-going research project. The focus of this paper is to assess the feasibility of developing a federated ERP system. Based on the proposed architecture of federated ERP systems [12], [13], [15], [16], [17] a certain complexity is given. Looking at the decision making process that can be found in small and medium size companies [9], [23] several decision criteria can be found which have an influence on the overall adoption of federated ERP systems. These directly influence the uncertainty of success when building such a system. In order to formalize the assessment, distinct criteria concerning the effort and effect are used. Different aspects regarding strategic planning [24] will be taken into consideration when formalizing the criteria catalogue. The rating of these criteria will be done based on data from the ERP industry [25], [26], [27], [28], [29] regarding Research & Development (R&D) cost, but also development time and profit. For the net present value calculation, a weighted average cost of capital is estimated using referential data from software companies. [30], [31], [32] The goal is to obtain a first assessment of the overall feasibility which is independent of a specific company which would try to develop such a system. Additionally, potential risks that influence the feasibility will be identified. The identified problems will not be analysed as part of this paper; further research is required to clarify these issues. The following figure summarizes the research method.

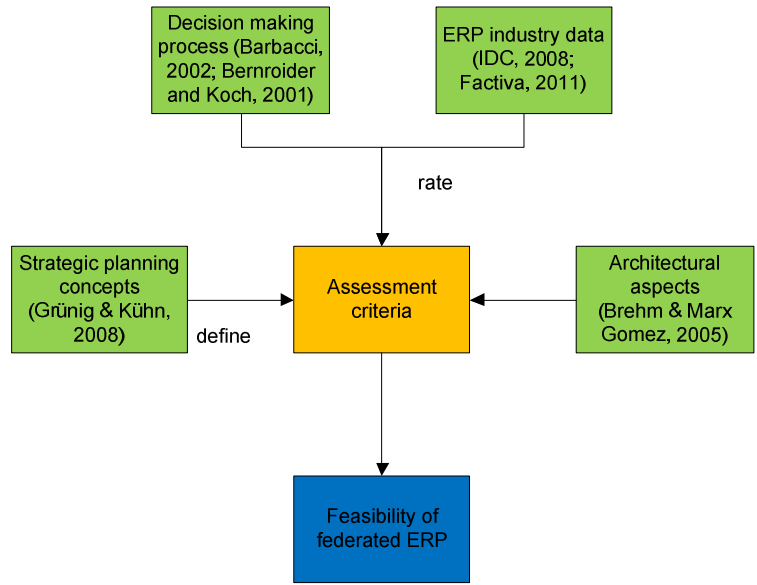

Figure 1. Research method

## 4. Federated ERP

Traditional ERP systems integrate various types of business applications into one system. These systems cover various different functions in the enterprise. Which functionality of the overall pack-age is used in a customer installation depends on the needs of the customer and is configured by either the implementation partner or ERP vendor. The installing, developing and maintenance of this system is very expensive. Small and medium size businesses are especially unable to cover the cost of such systems. Modern ERP systems consist of many software components which are related to each other and administered on a central application server. [12], [13] Due to the complexity of these systems, Brehm and Marx Gómez [10], [11], [12], [13] identified several problems. The main issues are the expensive customizations and the high end computer hardware requirements, which lead to high cost. One solution to the problems around traditional ERP systems is to develop a distributed ERP system where the system components are reachable and accessible over a network. This idea has been formulated as the concept of Federated ERP-System on the basis of Web-Services. [10], [11], [12], [13], [15], [22]

A federated ERP system (FERP system) is an ERP system which consists of system components that are distributed within a computer network. The overall functionality is provided by an ensemble of allied network nodes that all together appear as a single ERP system to the user. Different ERP system components can be developed by different vendors. [12], [13], [14]

This ensemble of component appears towards the user still as single ERP system. Physically, it consists of different independent elements which reside on different computers. This architecture allows an enterprise to access on-demand functionality (components) as services of other network members over a P2P network. [33] The architecture proposed by Brehm and Marx Gomez consists of several interconnected subsystems. The main objective is to integrate business components of different vendors, all of which need to follow the same standard, which will be described as XML schema documents. [16], [17]

## 5. ASSESSMENT CRITERIA

Various sources deal with the term feasibility study. [34], [35], [36] The purpose of a feasibility study is to determine if a business opportunity is possible and viable [34], [37] defines six dimensions of business viability: Market, technical, business model, management model, economic and financial model and exit strategy viability. [37] Various frameworks exist to assess feasibility of the different dimensions. The architecture trade-off analysis method, as proposed by Klein [38], is useful to assess a given architecture, and to evaluate the constraints given by the architecture. It is used to assess the feasibility when implementing an ERP project. [23] The cost benefit analysis method [23] is architecture centric and evaluates the benefits that a given architecture provides to a company using the system. Both methods are in the ERP selection phase and valuable to assess the feasibility of an ERP implementation project, but their application for business innovation is not given. [23] Grünig and Kühn [24] define different stages during the assessment of a strategy. During each phase different models are used. SWOT Analysis, Porters five forces model and the Boston Consulting Group portfolio matrix are some of them. [24] These models are covering the different viability dimensions. [24] Theses assessment models take the current market position and resources of a firm into consideration. Due to the fact, that an overall assessment of a federated ERP strategy should be done, no concrete market position or resources exist. Saaty [39] is proposing an analytical hierarchal process (AHP) for decision making. This model is useful to structure complex decision situations [39] and can be used during the development phase of a federated ERP system. Hence for an overall assessment, not enough information is available to build the hierarchical graph of the decision. So in order to retrieve the overall feasibility assessment, a different approach is taken. To assess the overall feasibility of the development of a federated ERP system, the effect of the strategy and the effort will be used. To provide a framework to obtain the effect and the effort of a strategy, within each category sub criteria are applied, following an analytical hierarchical approach. The effect / effort are then calculated as the average of the sub criteria. All sub criteria are weighted equally. In order to allow an objective rating, the domain of the criteria has been limited. Each criterion has only three distinct values. The different criteria will be explained in the following section:

### 5.1. Effect

#### 5.1.1. Impact time horizon

This criterion assesses how long it will take, before a developed system will be adopted within the market and start to show the anticipated effect. A distinction between long term and short term can be done. Looking at the overall technology adoption lifecycle for new products, a distinction be-tween innovators, early adopters, early majority, late majority and laggards can be found. The technology adoption lifecycle describes when customers start switching to a new technology. [40] So for the criteria values, 2 years is used as the shortest impact time horizon, taking the development time and the adoption rate of the product into consideration. The following distinct values are used:

1.     Long (Greater than 4 years)
2.     Medium (Within 2 and 4 years)
3.     Short (Less than 2 years)

#### 5.1.2. Expected value gain

This criterion should assess the potential of a federated ERP system. How much customers will benefit once the system has been developed successfully. In order to assess this, an average of the revenues from some of the major traditional ERP vendors has been taken as the base line. Looking at the ERP industry [25], [26], [27], [28], [29] an average revenue of 100 million $ for the new system can be estimated. This is used as the baseline for this criterion.

1.  Low (Less than 100 million $)
2.  Medium (Between 100 million and 200 million $)
3.  High (More than 200 million $)

### 5.1.3. Uncertainty

In order to assess certainty of the value gain, this ratio is used. This criterion should assess the risk attached to a project to develop a federated ERP system, and also the uncertainty of the outcome of the project. It can be distinguished further into the following distinct values:

1.  High (Outcome of the project is uncertain)
2.  Medium
3.  Low (No big uncertainties)

## 5.2. Effort

### 5.2.1. Development time frame

In which time frame can the system be developed? This criterion assesses how fast the system can be put in place. Based on the average release cycle of an ERP product, 2 years is used as the baseline. The following distinct values are used:

1.  Short (Less than 2 years)
2.  Medium (Within 2 and 4 years)
3.  Long (Greater than 4 years)

### 5.2.2. Scope

This criterion classifies whether the system can be developed by only applying internal resources, only external resources of a system implementer, or both. So for example, which parties need to be involved to put the system in place? Due to the fact, that managing internal factors is less complex then external ones, the following rating is used.

1.  Internal factors only
2.  External factors only
3.  Combination of internal and external factors

### 5.2.3. Resources

Which resources are required to develop the system and can these resources be accessed. In order to get a comparison between what it currently requires to develop a traditional ERP system, the average R&D cost of major ERP vendors has been taken as the baseline. On average the major ERP vendors [41], [42], [43] spend 250 million $ per product line on R&D. [25], [26], [27], [28], [29] This is used as the baseline for the development of a federated ERP platform. The following ratios are used:

1.  Low (Less than 250 million $)
2.  Medium (Between 250 million and 500 million $)
3.  High (More than 500 million $)

### 5.2.4. Complexity

How complex is the development of the system? Which dependencies exist between the anticipated involved components and parties? Looking at project management literature, a project which spans multiple companies and has multiple activities which are impacting different parts of the company is rated as a complex. [39] The following distinct values are applied:

1.    Low (Few teams involved, easy to manage)
2.    Medium
3.    High (Complex network of parties involved)

# 6. ASSESSMENT

In order to obtain an assessment of the feasibility for the development of a federated ERP system, the hierarchical criteria defined in previous chapter are applied. In this section the assessment and reasoning of how the rating is obtained is discussed. In order to have some referential points regarding development time and cost, the average values [25], [26], [27], [28], [29] from the top 3 vendors from each of the three ERP market segments, small, medium size and large enterprises, which are identified by Pang [41], [42], [43], will be used. The overall assessment is shown in figure 2.

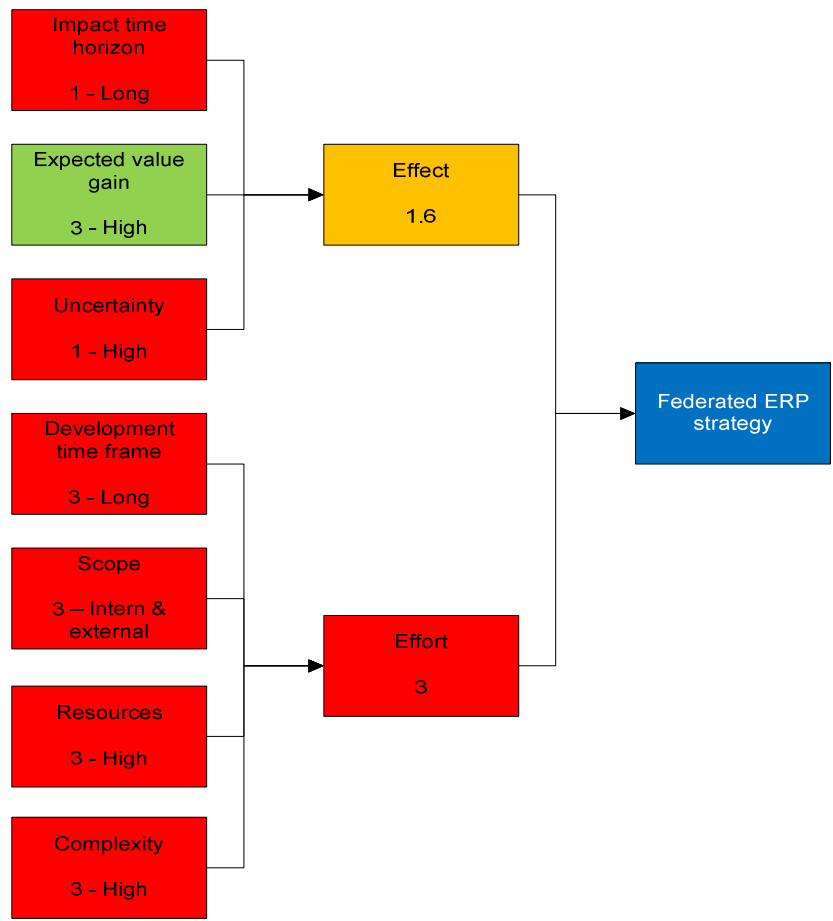

Figure 2. Feasibility assessment federated ERP system

## 6.1. Effect

### 6.1.1. Impact time horizon = 1 - Long

The new concept would change the ERP market drastically. The whole business model would need to change. The customer reaction on new technology can be described using the technology adoption life cycle from Moore [40]. Some early adopters might start using the technology when it comes to the market, but due to the fact that the ERP system is one of the core components within the corporate IT structure, it is more likely that a majority of the customers will wait until the technology is more mature. This leads to a rating of this criterion with 1 – Long.

### 6.1.2. Expected value gain = 3 - High

This strategy is the most risky alternative, because the adaption of federated ERP systems is still unclear. Marx Gómez and Brehm [12], [13], [15], [16], [17] also describe that the solutions that should be provided must be at a lower price, because this is one of the main incentives for customers to switch to federated ERP systems. Some of the web services that are provided are now already on an open source basis, so the main revenue would need to come from the basic platform, which is comparable to products like Visual Studio or IBM Websphere, so therefore this criteria is rated as 3 - high.

### 6.1.3. Uncertainty = 1 - High

As already mentioned, the adoption of such a game changing platform is unclear so therefore this option has a very high uncertainty. It might end up being a complete failure if customers do not fully trust web based systems. Looking at the ERP selection process in SME, the market position of the ERP vendor plays an important role, as does the vendor's support. [9] Introducing a fundamental new concept needs to mitigate these expectations. Currently, the customer adoption rate for federated ERP systems has not been evaluated. Trusting multiple different vendors and the liability problem needs to be investigated further. So therefore the uncertainty is rated with 1- High.

## 6.2. Effort

### 6.2.1. Development time frame = 3 - Long

Building a completely new platform takes more time and involves more dependencies and testing compared to releasing a new version or product on an existing platform. Taking into consideration that there are no federated ERP systems on the market right now and that a significant part of the system is defining the communication standards [12], [13], [16], [17] the anticipated discoveries can be rated as high. Therefore, it cannot fit within one release cycle, which leads to a rating of 3 – Long.

### 6.2.2. Scope = 3 – Internal & External

This solution describes a lot of research and collaboration with universities, customers and partners and involves both internal and external parties. Brehm and Marx Gomez point out that communication between different components needs to be standardized [12], [13], which would require coordination between the different parties. Looking at the time it takes to develop an industry wide new standard, this is one of the major obstacles to be overcome for a successful development.

### 6.2.3. Resources = 3 - High

The development cost of a new FERP platform is comparable to the development of a general development platform (e.g. IBM Websphere) where various different vendors would need to adhere to the same standard. Various different specialized resources from both inside and outside the company would be required, which leads to a rating of 3 – High.

### 6.2.4. Complexity = 3 - High

As already mentioned, the different involved parties that need to be coordinated result in a complex project setup. [43] Additionally, legal aspects of the project regarding the liability of the framework and the ensemble of various different questions need to be clarified. All these issues together lead to a rating of 3 – High in regard to the complexity. Many currently unknown variables need to be managed. So here also, additional research would be required to frame the project scope better.

## 7. NET PRESENT VALUE

Using the in the previous chapter mentioned assessment of the effect and the effort a net present value calculation can be done. In order to use an appropriate weighted average cost of capital, referential data from the software industry is used [30], [31], [32] to estimate the weighted average, which leads to a value of 10 % which will be used for the net present value calculation. The net present value is calculated following Brealey, Myers and Allen. [45] In order to develop a federated ERP system, an initial investment of 500 million dollar over two years is estimated. During this initial phase, the industry wide standards which are necessary for a successful development of a federated ERP system need to be set. Additional, the platform for the system needs to be developed. Therefore, a higher investment during this period is estimated. After this initial period, 250 million dollar is estimated to stabilize the developed system. This is following the average R&D spending of the bigger ERP vendors. [25], [26], [27], [28], [29] Regarding the estimated return, the current revenue of successful ERP vendors is used as a baseline. [25], [26], [27], [28], [29] Using these values, the net present value can be calculated as followed:

Table 1. Net present value

| Year | 0 | 1 | 2 | 3 | 4 | 5 | 6 | 7 | 8 | 9 |
|------|------|------|------|------|-----|-----|-----|-----|-----|-----|
| Investment | -500 | -500 | -250 | -250 | | | | | | |
| PV (Investment) | -500 | -455 | -207 | -188 | | | | | | |
| Return | | | | | 600 | 600 | 600 | 600 | 600 | 600 |
| PV (Return) | | | | | 410 | 373 | 339 | 308 | 280 | 254 |
| Total PV (Invest) | -1349 | | | | | | | | | |
| Total PV (Return) | 1.963 | | | | | | | | | |
| Total | 614 | | | | | | | | | |

As shown in "Table 1", the development of a federated ERP system does show a positive net present value. However, in order to obtain a detailed investment and return on investment, the mentioned uncertainties regarding the adoption of federated ERP systems within the customer base of ERP users' needs to be assessed further. Taking the market size of the ERP market and the growth rate into consideration potentially even bigger revenue could be expected. [41], [42] Looking at different weighted average cost of capital values, the development of such a system provides a higher return, if a lower weighted average cost of capital is assumed, which are currently reflected by bigger software companies. [30], [31], [32]

## 8. CONCLUSION

Several problems arise when implementing an ERP system in a company. The various different process optimization strategies may mitigate some of these problems, but does not address the root cause, the huge complexity of ERP systems. The different approach of federated ERP does provide a different approach. This paper assesses the feasibility of developing such a system. The overall feasibility of developing a federated ERP system shows high potential effect, but also high effort attached to it. So it would require a large amount of money and time to develop a standardized new federated ERP platform. For a company, the first mover advantage, as well as setting the industry standard could be an additional incentive to move into the area of federated ERP. A detailed net present value calculation, which is made for a concrete company which would think about implementing a federated ERP system, needs to be done. Small and medium size companies would benefit from federated ERP systems due to lower complexity. The low saturation of ERP installations in this market would provide additional opportunities. The main obstacles identified are the achievement of an industry wide standard and the change in customer behaviour with regard to ERP. Further research needs to be done, especially with regard to customer behaviour. It is unclear yet, if companies would put their trust in several small components, where the liability aspect is more difficult compared to a traditional ERP system. All in all, federated ERP provides attractive opportunities and interesting problems to be solved.

**Authors**

Michael Gall is a researcher at the Research Group for Industrial Software (INSO), Technical University Vienna. His research interest included ERP architectures, implementation and refactoring. In the past 10 years he had been working as a software developer and consultant for ERP systems.

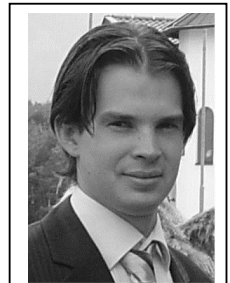

Thomas Grechenig is Head of the research group for Industrial Software (INSO). He is author of scientific work in the fields of software engineering, software quality management and project management. He has 20 years of practical experience. Current industrial research application fields are web-engineering, web-banking and M-commerce.

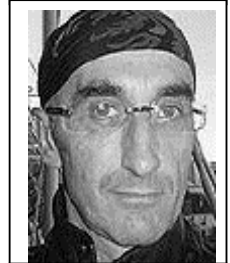

Mogens Bjerre is associate professor at Copenhagen Business School. He is affiliated with the Research Group for Consumer Behaviour. His primary research areas are Consumer behaviour and key account management. He is teaching MBA and executive MBA courses at Copenhagen Business School.

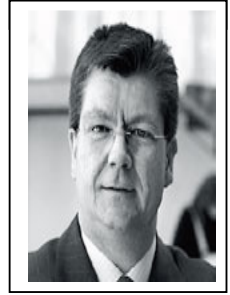